\documentclass[twocolumn,conference,letterpaper]{IEEEtran}
\usepackage[left=0.75in,right=0.75in,top=1in,bottom=0.75in]{geometry}

\usepackage{cite}
\usepackage{amssymb,amsmath,latexsym,amsfonts,amsthm,mathtools}

\usepackage{graphicx}
\usepackage{float,subcaption}
\usepackage{textcomp}
\usepackage{xcolor}
\definecolor{darkpastelpurple}{rgb}{0.59, 0.44, 0.84}
\usepackage{hyperref}
\hypersetup{colorlinks, breaklinks, citecolor=blue, linkcolor=blue, urlcolor=blue}
\usepackage[nameinlink]{cleveref}
\Crefformat{figure}{#2Fig.~#1#3}
\Crefmultiformat{figure}{Figs.~#2#1#3}{ and~#2#1#3}{, #2#1#3}{ and~#2#1#3}

\usepackage{tikz}
\usetikzlibrary{automata, shapes, arrows, calc, arrows.meta, fit, positioning}

\theoremstyle{plain}
\newtheorem{theorem}{Theorem}
\newtheorem{lemma}{Lemma}

\newtheorem*{problem*}{Problem}
\theoremstyle{remark}

\newtheorem{assumption}{Assumption}

\theoremstyle{definition}

\newcommand{\sign}{\mathrm{sign}}
\usepackage{mathrsfs}

\usepackage{newtxmath}
\usepackage{booktabs}
\usepackage{duckuments}

\IEEEoverridecommandlockouts

\begin{document}
	\title{Nonlinear Cooperative Salvo Guidance with Seeker-Limited Interceptors}
	\author{Lohitvel Gopikannan, Shashi Ranjan Kumar,~\IEEEmembership{Senior Member, IEEE}, and Abhinav Sinha,~\IEEEmembership{Senior Member,~IEEE}
		\thanks{L. Gopikannan and S. R. Kumar are with the Intelligent Systems \& Control (ISaC) Lab, Department of Aerospace Engineering, Indian Institute of Technology Bombay, Mumbai 400076, India. (e-mails: 24m0023@iitb.ac.in, srk@aero.iitb.ac.in). A. Sinha is with the Guidance, Autonomy, Learning, and Control for Intelligent Systems (GALACxIS) Lab, Department of Aerospace Engineering and Engineering Mechanics, University of Cincinnati, OH 45221, USA. (e-mail: abhinav.sinha@uc.edu).}
	}

	\maketitle
	\thispagestyle{empty}
	
	\begin{abstract}
		This paper presents a cooperative guidance strategy for the simultaneous interception of a constant-velocity, non-maneuvering target, addressing the realistic scenario where only a subset of interceptors are equipped with onboard seekers. To overcome the resulting heterogeneity in target observability, a fixed-time distributed observer is employed, enabling seeker-less interceptors to estimate the target state using information from seeker-equipped agents and local neighbors over a directed communication topology. Departing from conventional strategies that approximate time-to-go via linearization or small-angle assumptions, the proposed approach leverages deviated pursuit guidance where the time-to-go expression is exact for such a target. Moreover, a higher-order sliding mode consensus protocol is utilized to establish time-to-go consensus within a finite time. The effectiveness of the proposed guidance and estimation architecture is demonstrated through simulations.
	\end{abstract}
	
	
	\section{Introduction}\label{sec:introduction}
    Modern intercept guidance systems are designed to ensure precise target interception while also meeting terminal performance criteria such as fixed impact time and angle of impact. Although these objectives have been effectively addressed in single-interceptor engagement, practical challenges such as constrained sensor coverage, limited reaction times, and actuator saturation can significantly degrade interception reliability in operational platforms. Salvo guidance offers a more robust alternative by enabling multiple interceptors to engage the target collaboratively, thereby improving interception probability, redundancy, and overall system efficiency.
    
    Salvo strategies are typically classified as non-cooperative, where interceptors operate independently with fixed impact times, and cooperative, which allows real-time coordination. The former may not be reliable in ensuring simultaneous interception, primarily due to the absence of inter-interceptor communication, which prevents the correction of discrepancies in impact time estimates arising during flight. Early developments aimed at achieving simultaneous interception of stationary targets can be found in \cite{p10}, where multiple interceptors were guided to engage a common target at a predetermined impact time. To realize this objective, integral sliding mode control (SMC) techniques were proposed in \cite{p12} for regulating the time of impact. Finite-time coordination was further explored in \cite{p13} through a guidance strategy based on state-dependent Riccati equations. To address the shortcomings of non-cooperative salvo, a series of cooperative salvo strategies were developed in works such as \cite{p12},\cite{p6},\cite{p9},\cite{p14} and \cite{p15} where interceptors exchange time-to-go information in real time and achieve synchronization through distributed communication protocols. Individual impact time control was utilized in \cite{p17} to achieve salvo interception, and this approach was further improved by integrating a predicted interception point (PIP) component to intercept moving targets, as in \cite{p11}. A similar strategy using a leader–follower approach and PN guidance-based time-to-go estimate was proposed in \cite{p6}, where super-twisting sliding mode control was leveraged to ensure finite-time convergence of the impact time error under large heading angle errors. Note that PIP is an approximation, and consequently, several other strategies for intercepting moving targets have since been proposed.

    As opposed to the variants of PN guidance, time-constrained interception under the deviated pursuit scheme has been put forth in \cite{p4,p16,11018241}. Unlike PN guidance-based impact time control strategies that rely on time-to-go estimates, deviated pursuit guidance favors an exact analytical formulation of time-to-go against constant velocity targets, thereby enhancing both the accuracy and robustness of the guidance strategy by modulating the deviation angle to enable precise time-constrained interception. 

    It is worth noting that, typically, cooperative salvo guidance strategies in the contemporary literature (as seen in the aforementioned works and references therein) require full information about the target to synchronize their impact time values. From an autonomy-oriented systems design perspective, equipping all interceptors with onboard seekers is often infeasible in large-scale engagements due to constraints on cost, weight, and energy consumption. A scalable and autonomous architecture delegates sensing capabilities to a subset of agents, while the remaining interceptors operate seeker-less by autonomously estimating target states through distributed information fusion and local coordination. This enables decentralized decision-making and synchronized time-constrained interception without full sensing redundancy. Building on these considerations, this work investigates a cooperative salvo guidance scenario characterized by heterogeneous access to target state information, where only a subset of interceptors possesses onboard sensing capabilities. The remaining agents employ a fixed-time distributed observer to estimate the necessary relative engagement variables to facilitate cooperative salvo despite sensing asymmetries.

    \section{Preliminaries and Problem Formulation}
\begin{figure}[h!]
        \centering
        \includegraphics[width=.7\linewidth]{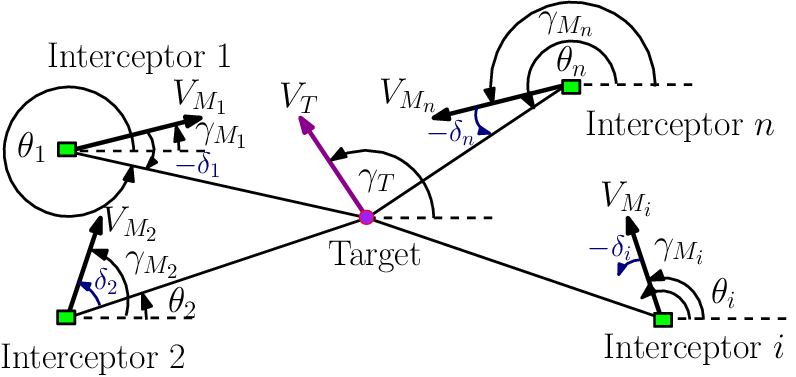}
        \caption{Cooperative multiagent engagement.}
        \label{fig:enggeo}
    \end{figure}

    Consider a multiagent pursuit scenario involving $n$ interceptors and a single target moving with constant velocity $V_T$, as shown in \Cref{fig:enggeo}. The speeds of the interceptors are $V_{M_{i}}$ corresponding to the $i$\textsuperscript{th} interceptor, whereas the relative range and line-of-sight (LOS) angle w.r.t the target are $r_i$ and $\theta_i$, respectively. The flight path angle of the target is $\gamma_T$, while that of the $i$\textsuperscript{th} interceptor is $\gamma_{M_i}$. The $i$\textsuperscript{th} interceptor is steered using its lateral acceleration $a_{M_i}$, whereas $\delta_i$ is its deviation angle.
    Accordingly, the equations of relative motion are expressed as
    \begin{subequations}\label{eq:enggeo}
   \begin{align}
V_{r_i} = \dot{r}_i &= V_T \cos(\gamma_T - \theta_i) - V_{M_i} \cos\delta_i, \\
V_{\theta_i} =r_i\dot{\theta}_i&= V_T \sin(\gamma_T - \theta_i) - V_{M_i} \sin\delta_i,
\end{align}
\end{subequations}
where $\delta_i=\gamma_{M_i}-\theta_i$, while $V_{r_i}$ and $V_{\theta_i}$ are the relative velocity components along and across the LOS.
\begin{assumption}\label{asm:seeker}
Only a subset of interceptors is equipped with onboard seekers, enabling them to directly measure the target state, while the remaining interceptors rely on inter-agent communication over a directed graph $\mathcal{G}$ to estimate the target state via local information exchange with neighboring agents.
\end{assumption}    
\Cref{asm:seeker} is concerned with a heterogeneous sensing architecture, which is central to enabling the joint estimation and guidance approach presented in our paper, allowing the entire interceptor team to coordinate effectively despite partial observability. This allows us to formulate the main problem addressed in our paper.
\begin{problem*}
    Subject to \Cref{asm:seeker}, design a cooperative guidance strategy to ensure a simultaneous interception of a target moving with constant velocity.
\end{problem*}
Designing cooperative salvo guidance strategies may entail control over interceptors' time-to-go. Thus, we propose to utilize deviated pursuit guidance as a baseline, which offers an exact expression of time-to-go\cite{p4}
\begin{equation}\label{eq:tgo}
    t_{\mathrm{go}_i} = \dfrac{r_i\left(V_{r_i}+2V_{M_i}\cos\delta_i-V_{\theta_i}\tan\delta_i\right)}{V_{M_i}^2-V_T^2}
\end{equation}
against a target moving with constant velocity, written for the $i$\textsuperscript{th} interceptor. Note that \eqref{eq:tgo} requires engagement variables such as relative range and LOS angles w.r.t. to the target, which may not be available to seeker-less interceptors, and hence, the cooperative guidance command for the $i$\textsuperscript{th} interceptor must be designed on the estimates of such variables, which form the information about the target.

To this end, let the \emph{sensing} topology be modeled by a directed graph $\mathcal{G}=\{\mathcal{V},\mathcal{E}\}$, where the node set $\mathcal{V}=\{v_0, v_1, \ldots, v_n\}$ comprises a single target agent $v_0$ and $n$ interceptors $v_1$ through $v_n$. The directed edge set $\mathcal{E}\subseteq \mathcal{V}\times \mathcal{V}$ captures the communication topology, where an edge $e_{ij}\in \mathcal{E}$ denotes a directed information link from agent $v_i$ to agent $v_j$. For each node $v_i\in \mathcal{V}$, the in-neighbor set is defined as $\mathcal{N}_i=\{v_j\in\mathcal{V}~|~e_{ji}\in\mathcal{E}\}$, representing all agents from which $v_i$ receives information. An agent is classified as a leader if $\mathcal{N}_i=\emptyset$, and a follower otherwise. In this context, the target $v_0$ serves as the unique leader (i.e., it transmits but does not receive information), while all interceptor agents act as followers. The communication structure is encoded in the adjacency matrix $\mathcal{A}=[a_{ij}]\in\mathbb{R}^{(n+1)\times (n+1)}$, with 
\begin{equation*}
    a_{ij}=\begin{cases}
        1,&\text{if}~ e_{ji}\in \mathcal{E}\\
        0, & \text{otherwise}
    \end{cases}\quad\text{and}~a_{ii}=0.
\end{equation*}
The in-degree matrix is defined as $\mathcal{D}=\text{diag}\{d_0,d_1,\ldots,d_n\}$, where each diagonal entry $d_i=\sum_{j=0}^{n}a_{ij}$ represents the number of incoming edges to node $v_i$. The associated graph Laplacian is given by $\mathcal{L}=\mathcal{D}-\mathcal{A}$, and the Kronecker product $\otimes$ is used to construct higher-dimensional system dynamics from scalar graph structures.
\begin{assumption}[\cite{p8}] 
The directed graph $\mathcal{G}$ is assumed to contain a directed spanning tree rooted at the target $v_0$, and then the associated Laplacian matrix can be expressed as 
\begin{equation}
    \mathcal{L} = 
    \begin{bmatrix}
        0 & \mathbf{0}_{1\times n} \\
        \mathcal{L}_{TI} & \mathcal{L}_{II}
    \end{bmatrix},
\end{equation}
where $\mathcal{L}_{TI} \in \mathbb{R}^{n \times 1}$ captures the influence of leader node (target) on each follower (interceptors), and $\mathcal{L}_{II} \in \mathbb{R}^{n \times n}$ is the reduced Laplacian for the follower subgraph. 
\end{assumption}
\begin{lemma}
The eigenvalues of $\mathcal{L}_{II}$ are all positive such that $0 < \text{Re}\{\lambda_1\} \leq \text{Re}\{\lambda_2\} \leq \cdots \leq \text{Re}\{\lambda_n\}$.  
\end{lemma}
\begin{lemma}[\cite{p2}]\label{lem:ineq}
There exists a positive definite diagonal matrix $\bar{\mathcal{D}}\in\mathbb{R}^{n\times n}$ such that the symmetric part of the weighted Laplacian satisfies $\mathcal{L}_{II}^\top \bar{\mathcal{D}} + \bar{\mathcal{D}} \mathcal{L}_{II} \succ 0$. Let $\lambda_m$ denote the smallest eigenvalue of $\mathcal{L}_{II}^\top \bar{\mathcal{D}} + \bar{\mathcal{D}} \mathcal{L}_{II}$ and define a scaled diagonal matrix $\hat{\mathcal{D}} = \text{diag}(d_1, \ldots, d_n) =\dfrac{2\bar{\mathcal{D}}}{\lambda_m}$, then the inequality $\mathcal{L}_{II}^\top \hat{\mathcal{D}} + \hat{\mathcal{D}} \mathcal{L}_{II} \geq 2\mathbf{I}_n$ holds.
\end{lemma}
Let $\mathcal{G}^a = (\mathcal{V}^a, \mathcal{E}^a) $ denote the \emph{actuation graph}, representing the interaction topology used for coordinating guidance commands among interceptors. Here, $\mathcal{V}^a = \{v_1^a, \dots, v_n^a\}$ consists only of the interceptor nodes, and $\mathcal{E}^a \subseteq \mathcal{V}^a \times \mathcal{V}^a$ is the set of directed edges that need not coincide with $\mathcal{G}$. Unlike the sensing graph, the actuation graph is assumed to be \emph{leaderless}. The associated Laplacian matrix with $\mathcal{G}^a$ is given by $L_I$.
\begin{lemma}[\cite{p1}]\label{lem:ft}
Consider the dynamical system $\dot{x}(t) = g(t, x(t)), \quad x(0) = x_0 \in \mathbb{R}^n$, and suppose there exists a continuously differentiable function $V : \mathbb{R}^n \rightarrow \mathbb{R}_{\geq 0}$ such that $V(x)$  is positive definite, radially unbounded function and satisfies $V(x) = 0$ iff $x \in M \subset \mathbb{R}^n$. If the inequality $\dot{V}(x) \leq -(aV(x)^{m} + b V(x)^{n})^k
$ is satisfied by any solution  \( x(t) \)  for some constants \( a,b, m, n , k > 0 \), with \( mk < 1 \) and \( nk > 1 \), then the set \( M \) is globally fixed-time stable, and the settling time \( T(x_0) \) for any initial condition \( x_0 \in \mathbb{R}^n \) satisfies the upper bound $T(x_0) \leq \frac{1}{a^k (1 - mk)} + \frac{1}{b^k (nk - 1)}$.
\end{lemma}
\begin{lemma}\label{lem:gammineq}
Let \( \rho_1, \rho_2, \dots, \rho_n \in \mathbb{R} \), and \( p \in \mathbb{R}_+ \). If \( 0 < p \leq 1 \), \( \left( \sum_{i=1}^n |\rho_i| \right)^p \leq \sum_{i=1}^n |\rho_i|^p \leq n^{1 - p} \left( \sum_{i=1}^n |\rho_i| \right)^p \). If \( p > 1 \), then \( \sum_{i=1}^n |\rho_i|^p \leq \left( \sum_{i=1}^n |\rho_i| \right)^p \leq n^{p-1}  \sum_{i=1}^n |\rho_i|^p \).  
\end{lemma}

 \section{Joint Estimation and Guidance Framework}
    Recall that designing a cooperative salvo guidance may necessitate control over engagement duration and that \eqref{eq:tgo} consists of engagement variables typically not accessible to all the interceptors. Following \Cref{asm:seeker}, the seeker-less interceptors must estimate the relevant engagement variables. To this end, we consider the target states to be its position $(X_T,Y_T)$ and velocity $(\dot{X}_T,\dot{Y}_T)$ in the inertial frame, such that the target state dynamics can be written as
    \begin{align}
        \dot{\mathbf{z}} = \mathbf{A} \mathbf{z} \implies \begin{bmatrix}
            \dot{X}_T \\ \dot{Y}_T\\ \ddot{X}_T\\ \ddot{Y}_T
        \end{bmatrix}=
\begin{bmatrix}
0 & 0 & 1 & 0 \\
0 & 0 & 0 & 1 \\
0 & 0 & 0 & 0 \\
0 & 0 & 0 & 0
\end{bmatrix}\begin{bmatrix}
            X_T \\ Y_T\\ \dot{X}_T\\ \dot{Y}_T
        \end{bmatrix}.
    \end{align}
Let us denote the estimate of $\mathbf{z}$ as observed by the $i$\textsuperscript{th} interceptor as $\hat{\mathbf{z}}^i$ and let $\mathbf{\eta}^i$ be the inter-agent relative estimation error. Following \cite{p2}, the fixed-time distributed observer for each interceptor to estimate $\mathbf{z}$ is given by
\begin{align}
    \dot{\hat{\mathbf{z}}}^i =& \mathbf{A} \hat{\mathbf{z}}^i - k_1\eta^i-k_2\mathrm{sig}^\alpha \eta^i-k_3\mathrm{sig}^{\beta}\eta^i, \label{eq:Stateobserver}
\end{align}
where $\mathrm{sig}^{(\star)}(\cdot)=|(\cdot)|^{\star}\sign(\cdot)$, $k_1,k_2,k_3,\alpha,\beta$ are some design parameters satisfying $k_1, k_2, k_3 > 0$, $\alpha\in(0,1)$, $\beta>1$, and
\begin{align}
    \eta^i =& a_{i0} (\hat{\mathbf{z}}^i -{\mathbf{z}})+ \sum_{j=1}^n a_{ij} (\hat{\mathbf{z}}^i -\hat{\mathbf{z}}^j).
    \label{relative estimation error}
\end{align}
\begin{theorem}
Consider the cooperative target engagement described by \eqref{eq:enggeo}, where only a subset of interceptors has direct access to the target states. Let each interceptor implement a distributed observer over a directed sensing graph $\mathcal{G}$. Then, under the observer dynamics \eqref{eq:Stateobserver}, the $i$\textsuperscript{th} interceptor's local estimate converges to the 
true target states within a fixed-time $T_s$ regardless of their initial estimates, that is, 
\begin{align}
    \lim_{t \to T_s} (\hat{\mathbf{z}}^i(t)-\mathbf{z}(t)) &= 0.\label{observerstab}
\end{align}
\end{theorem}
\begin{proof}
We adopt vector-matrix notations hereafter to streamline the proof. Denote $\hat{\mathbf{z}} = [\hat{\mathbf{z}}^{1^\top}, \ldots, \hat{\mathbf{z}}^{n^\top}]^\top$ and $\eta = [\eta^{1^\top}, \ldots, \eta^{n^\top}]^\top$ and define $\delta^i= k_1 \eta^i + k_2 \text{sig}^\alpha \eta^i + k_3 \text{sig}^\beta\eta^i$ such that $\delta=[\delta^{1^\top},\ldots,\delta^{n^\top}]^\top$. Then, one may express \eqref{eq:Stateobserver} as 
\begin{equation}
\dot{\hat{\mathbf{z}}} = (\mathbf{I}_n \otimes \mathbf{A})\hat{\mathbf{z}} - (\mathbf{I}_n \otimes \mathbf{I}_{4})\delta.
\end{equation}
Define the local estimation error for the $i$\textsuperscript{th} interceptor as $\tilde{\mathbf{z}}^i = \hat{\mathbf{z}}^i - \mathbf{z}$ such that $\tilde{\mathbf{z}} = [\tilde{\mathbf{z}}^{1^\top}, \ldots, \tilde{\mathbf{z}}^{n^\top}]^\top$. Then,
\begin{equation}
\dot{\tilde{\mathbf{z}}} = (\mathbf{I}_n \otimes \mathbf{A})\tilde{\mathbf{z}} - (\mathbf{I}_n \otimes \mathbf{I}_{4})\delta.
\end{equation}
From \eqref{relative estimation error}, it follows that $\eta = (\mathcal{L}_{II} \otimes \mathbf{I}_{4}) \tilde{\mathbf{z}}$, whose dynamics becomes
\begin{equation}
\dot{\eta} = (\mathbf{I}_n \otimes \mathbf{A})\eta - (\mathcal{L}_{II} \otimes \mathbf{I}_{4})\delta.
\label{errorequation}
\end{equation}
Now, consider a Lyapunov function candidate
\begin{align}
V(\eta) = &\sum_{i=1}^n \left( \dfrac{k_2 d_i}{1 + \alpha} \left\|\eta^i\right\|_{1+\alpha}^{1+\alpha} 
+ \dfrac{k_3 d_i}{1 + \beta} \left\|\eta^i\right\|_{1+\beta}^{1+\beta} \right) \nonumber\\
&+ \dfrac{k_1 \eta^\top (\hat{\mathcal{D}} \otimes \mathbf{I}_4) \eta}{2},\label{Lyapunov}
\end{align}
whose time derivative is
\begin{align}
\dot{V}(\eta) 
= &\sum_{i=1}^n \left( k_2 d_i \, \text{sig}^\alpha\eta^i 
+ k_3 d_i \, \text{sig}^\beta\eta^i \right)^\top \dot{\eta}^i \nonumber\\
&+ k_1 \eta^\top (\hat{\mathcal{D}} \otimes \mathbf{I}_4) \dot{\eta}
\end{align}
and can be simplified to
\begin{equation}
 \dot{V}(\eta)  = \delta^\top (\hat{\mathcal{D}} \otimes \mathbf{A})\eta -\dfrac{\delta^\top \left[\left(\mathcal{L}_{II}^\top \hat{\mathcal{D}} + \hat{\mathcal{D}} \mathcal{L}_{II}\right) \otimes \mathbf{I}_4\right]\delta}{2}.
\label{v_dot_delta}
\end{equation}
Using Young's inequality and the results in \Cref{lem:ineq}, the expression in \eqref{v_dot_delta} can be further simplified to
\begin{align}
\dot{V}(\eta)
&\leq \frac{1}{2} \bigl( \| \hat{\mathcal{D}} \otimes \mathbf{A} \|^2 \|\eta\|^2 - \| \delta\|^2 \bigr).
\label{vdot13}
\end{align}




Since $k_1, k_2, k_3 > 0$, and the terms \( k_1 \eta^i \), \( k_2 \operatorname{sig}^\alpha\eta^i \), and \( k_3 \operatorname{sig}^\beta\eta^i \) are all sign-preserving with respect to \( \eta^i \), each component of the vector $\delta$ maintains the same sign as the corresponding component of $\eta^i$ ensuring directional consistency, which helps in deriving constructive lower bounds on \( \| \delta\|^2 \) in terms of $\|\eta\|$. Hence, one may write
\begin{align}
\|\delta\|^2 &= \sum_{i=1}^n \bigl\| k_1 \eta^i + k_2 \text{sig}^\alpha\eta^i + k_3 \text{sig}^\beta\eta^i \bigr\|^2 \nonumber\\
&= \sum_{i=1}^n\left( \bigl\| k_1^2 \|\eta^i\|^2 + k_2^2 \|\eta^i\|_{2\alpha}^{2\alpha} + k_3^2 \|\eta^i\|_{2\beta}^{2\beta} \right.\nonumber\\
&\left.+ 2k_1k_2\|\eta^i\|_{1+\alpha}^{1+\alpha}+2k_2k_3\|\eta^i\|_{1+\beta}^{1+\beta}+2k_1k_3\|\eta^i\|_{\alpha+\beta}^{\alpha+\beta}\right)\nonumber\\
&\geq \sum_{i=1}^n k_1^2 \|\eta^i\|^2 
+ \sum_{i=1}^n k_2^2 \|\eta^i\|_{2\alpha}^{2\alpha} 
+ \sum_{i=1}^n k_3^2 \|\eta^i\|_{2\beta}^{2\beta}.\label{delta}
\end{align}
Together with the constraints on the design parameters $\alpha,\beta$, it follows from the results in \Cref{lem:gammineq} that $\sum_{i=1}^n \|\eta^i\|_{2\alpha}^{2\alpha} \geq \bigl( \sum_{i=1}^n\|\eta^i\|^2 \bigr)^\alpha = \left(\|\eta\|^2\right)^\alpha$, and $\sum_{i=1}^n \|\eta^i\|_{2\beta}^{2\beta} \geq \sum_{i=1}^n \frac{1}{(4n)^{\beta - 1}} \bigl( \|\eta^i\|^2 \bigr)^\beta= \frac{1}{(4n)^{\beta - 1}} \bigl( \|\eta\|^2 \bigr)^\beta$.

Upon substituting from \eqref{delta} and the above relations in \eqref{vdot13}, $\dot{V}(\eta)$ can be expressed as


\begin{align}
\dot{V}(\eta) &\leq -\frac{1}{2} \bigl(k_1^2 - \| \hat{\mathcal{D}} \otimes \mathbf{A} \|^2\bigr) \|\eta\|^2 
- \frac{k_2^2}{2} \bigl(\|\eta\|^2\bigr)^\alpha \nonumber\\
&\quad - \frac{k_3^2}{2(4n)^{\beta - 1}} \bigl(\|\eta\|^2\bigr)^\beta\nonumber \\
&\leq -k \bigl( \|\eta\|^2 + \bigl(\|\eta\|^2\bigr)^\alpha + \bigl(\|\eta\|^2\bigr)^\beta \bigr),\label{vdot}
\end{align}

where, $k = \min \left\{ 
\frac{1}{2}({k}_1^2 - \|\hat{\mathcal{D}} \otimes \mathbf{A}\|^2),\ 
\frac{k_2^2}{2},\ 
\frac{{k}_3^2}{2(4n)^{\beta -1}} 
\right\} > 0$. Notice the terms in \eqref{Lyapunov} for which the relations
\begin{align}
    \sum_{i=1}^{n} \frac{k_2 d_i}{1+\alpha} \|\eta^i\|^{1+\alpha} \leq &  \frac{k_2 d_{\max}(4n)^{\frac{1 - \alpha}{2}}}{1 + \alpha}   \bigl(\|\eta\|^2\bigr)^{\frac{1+\alpha}{2}},\label{k2inqual}\\
    \sum_{i=1}^{n} \frac{k_3 d_i}{1+\beta} \|\eta^i\|^{1+\beta} \leq & \frac{k_3 d_{\max}}{1 + \beta} \bigl(\|\eta\|^2\bigr)^{\frac{1+\beta}{2}},\label{k3inequal}
\end{align}
where $d_{\max} = \max\{d_1, \ldots, d_n\}$, also follow from \Cref{lem:gammineq} since $\alpha\in(0,1)$ and $\beta>1$.

It follows from the relations \eqref{k2inqual}-\eqref{k3inequal}, $\eta^\top (\hat{\mathcal{D}} \otimes \mathbf{I}_{4}) \eta \leq \lambda_{\max}(\hat{\mathcal{D}}) \|\eta\|^2$, the results in \Cref{lem:gammineq},   and since $0 < \frac{2\alpha}{1+\alpha} < 1$, that
\begin{equation}
V^{\frac{2\alpha}{1+\alpha}} \leq c_\alpha \left( (\|\eta\|^2)^{\frac{2\alpha}{1+\alpha}} + (\|\eta\|^2)^{\alpha} + (\|\eta\|^2)^{\frac{\alpha(1+\beta)}{1+\alpha}} \right),
\end{equation}
such that $c_\alpha = \max \left\{ \left( \frac{k_1 d_{\max}}{2} \right)^{\frac{2\alpha}{1+\alpha}},
\left[\frac{k_2 d_{\max}(4n)^{\frac{1 - \alpha}{2}}}{1 + \alpha} \right]^{\frac{2\alpha}{1+\alpha}}\right.,$ $\left. 
\left[\frac{k_3 d_{\max}}{1 + \beta}\right]^{\frac{2\alpha}{1+\alpha}} \right\}$. As  \( \alpha < \frac{2\alpha}{1+\alpha} < 1 \), $\alpha<\frac{\alpha(1+\beta)}{1+\alpha}<\beta$, it follows that
\begin{align}
(\|\eta\|^2)^{\frac{2\alpha}{1+\alpha}} \leq \|\eta\|^2 + (\|\eta\|^2)^{\alpha},\\
(\|\eta\|^2)^{\frac{\alpha(1+\beta)}{1+\alpha}} \leq (\|\eta\|^2)^{\alpha}+ (\|\eta\|^2)^{\beta},
\end{align}
leading to 
\begin{equation}
V^{\frac{2\alpha}{1+\alpha}} \leq c_\alpha \left( \|\eta\|^2 + 3(\|\eta\|^2)^{\alpha} + (\|\eta\|^2)^{\beta} \right).
\label{calpha}
\end{equation}
Along similar lines, one may obtain
\begin{equation}
V^{\frac{2\beta}{1+\beta}} \leq c_\beta \left( (\|\eta\|^2)^{\beta}+(\|\eta\|^2)^{\frac{2\beta}{1+\beta}} + (\|\eta\|^2)^{\frac{\beta(1+\alpha)}{1+\beta}}\right),
\end{equation}
where $c_\beta = 3^{\frac{\beta - 1}{\beta + 1}}\max \Bigg\{ 
\left( \frac{k_1 d_{\max}}{2} \right)^{\frac{2\beta}{1+\beta}},\ 
\left[\frac{k_2 d_{\max}(4n)^{\frac{1 - \alpha}{2}}}{1 + \alpha} \right]^{\frac{2\beta}{1+\beta}},\allowbreak
\left[\frac{k_3 d_{\max}}{1 + \beta}\right]^{\frac{2\beta}{1+\beta}} 
\Bigg\}$ using the results in \Cref{lem:gammineq} and the relations  $\beta>1,1<\frac{2\beta}{1+\beta}<\beta$, and $\alpha<\frac{\beta(1+\alpha)}{(1+\beta)}<\beta$. Moreover, one also has
\begin{align}
(\|\eta\|^2)^{\frac{2\beta}{1+\beta}} \leq& \|\eta\|^2 + (\|\eta\|^2)^{\beta},\\
(\|\eta\|^2)^{\frac{\beta(1+\alpha)}{1+\beta}} \leq  &(\|\eta\|^2)^{\beta} + (\|\eta\|^2)^{\alpha}.
\end{align}
Therefore, one may write
\begin{equation}
V^{\frac{2\beta}{1+\beta}} \leq c_\beta \left( \|\eta\|^2 + 3(\|\eta\|^2)^{\beta} + (\|\eta\|^2)^{\alpha} \right).
\label{vebta}
\end{equation}

Combining \eqref{calpha} and \eqref{vebta}, we get,
\begin{equation}
\frac{1}{4c_\alpha} V^{\frac{2\alpha}{1+\alpha}} + \frac{1}{4c_\beta} V^{\frac{2\beta}{1+\beta}} 
\leq ( \|\eta\|^2 + \bigl(\|\eta\|^2\bigr)^\alpha + \bigl(\|\eta\|^2\bigr)^\beta \bigr),
\label{vterms}
\end{equation}
which can be used to simplify $\dot{V}(\eta)$ to
\begin{equation}\label{eq:Vdotfinal}
\dot{V}(\eta) \leq \frac{k}{4}\left(-\frac{V^{\frac{2\alpha}{1+\alpha}}}{c_\alpha}  
- \frac{V^{\frac{2\beta}{{1+\beta}}}}{c_\beta}\right)
\end{equation}
It follows from \Cref{lem:ft} that the system in \eqref{eq:Vdotfinal} is fixed-time stable with the upper bound on settling time given as
\begin{align}
    T_s \leq \frac{4c_\alpha (1 + \alpha)}{k (1 - \alpha)} + \frac{4c_\beta (1 + \beta)}{k(\beta - 1)}.
\end{align}
This essentially means that $\lim_{t\to T_s} \eta = 0$. Recall that $\eta= (\mathcal{L}_{II}\otimes \mathbf{I}_{4})\mathbf{\tilde z}$, and since $\mathcal{L}_{II}$ is nonsingular, it follows that $\lim_{t\to T_s} \mathbf{\tilde z} = 0$. Equivalently, $\lim_{t \to T_s} (\hat{\mathbf{z}}^i(t)-\mathbf{z}(t))=0$ and $\hat{\mathbf{z}}^i(t)-\mathbf{z}(t) =0,~\forall~t\geq T_s$. This completes the proof.

\end{proof}
The estimated engagament variables, namely, the range $\hat{r}_i$ and the LOS angle $\hat{\theta}_i$ as computed by the $i$\textsuperscript{th} interceptor are given by
\begin{align}
    \hat{r}_i=& \sqrt{(\hat X_{T_i}-X_{I_i})^2+(\hat Y_{T_i}-Y_{I_i})^2}\\
    \hat{\theta}_i=& \arctan\frac{\hat Y_{T_i}-Y_{I_i}}{\hat X_{T_i}-X_{I_i}},
\end{align}
whereas the target's heading angle is determined via
\begin{equation}
    \hat \gamma_{T_i}= \arctan\frac{\dot{Y}_{T_i}}{\dot{X}_{T_i}}.
\end{equation}
For guidance design, the equations of relative motion are now given in terms of the estimated engagement variables, that is,
\begin{subequations}\label{eq:enggeohat}
    \begin{align}
\hat V_{r_i} &= \hat V_{T_i} \cos(\hat \gamma_{T_i} - \hat \theta_i) - V_{M_i} \cos \hat \delta_i, \\
\hat V_{\theta_i} &= \hat V_{T_i} \sin(\hat \gamma_{T_i} - \hat \theta_i) - V_{M_i} \sin \hat \delta_i,
\end{align}
\end{subequations}
whereas the time-to-go \eqref{eq:tgo} modifies to
\begin{equation}\label{eq:tgohat}
    t_{\mathrm{go}_i} = \dfrac{\hat{r}_i\left(\hat{V}_{r_i}+2V_{M_i}\cos\hat{\delta}_i-\hat{V}_{\theta_i}\tan\hat{\delta}_i\right)}{V_{M_i}^2-\hat{V}_{T_i}^2}.
\end{equation}
\begin{lemma}
    The dynamics of the time-to-go \eqref{eq:tgohat} for the $i$\textsuperscript{th} interceptor, based on the estimated engagement variables, has a relative degree of one with respect to its lateral acceleration. 
\end{lemma}
\begin{proof}
On differentiating \eqref{eq:tgohat} with respect to time and using \eqref{eq:enggeohat}, one may obtain
    \begin{equation}
    \dot{t}_{\mathrm{go}_i} = -1 + \frac{\hat{V}_{\theta_i}^2 \sec^2 \hat{\delta}_i}{V_{M_i}^2-\hat{V}_{T_i}^2} - \frac{\hat{r}_i \hat{V}_{\theta_i} \sec^2 \hat{\delta}_i}{V_{M_i} \left(V_{M_i}^2-\hat{V}_{T_i}^2\right)} a_{M_i}
    \label{tgodynamics}
\end{equation}  
indicating that $a_{M_i}$ appears in the first derivative of ${t}_{\mathrm{go}_i}$.
\end{proof}
Let the error in the common time of simultaneous interception, for the $i$\textsuperscript{th} interceptor, be defined as
\begin{align}
    e_i = {t}_{\mathrm{go}_i} - {t}_{\mathrm{go}_d},\label{eq:ei}
\end{align}
where the desired time-to-go ${t}_{\mathrm{go}_d}$ is not required to be explicitly known but the interceptors will decide on a common value cooperatively as the engagement proceeds. Note that the distributed guidance commands must ensure consensus in the interceptors' time-to-go values, which is equivalent to saying that there must be an agreement in the error variable \eqref{eq:ei} since ${t}_{\mathrm{go}_i}-{t}_{\mathrm{go}_j}=e_i-e_j$ for any pair of interceptors $i,j$. Additionally, this error must also decrease to zero eventually to guarantee a cooperative salvo. To this end, we present the guidance command for the $i$\textsuperscript{th} interceptor in the next theorem.
\begin{theorem}
    Consider the cooperative target engagement described by \eqref{eq:enggeo}, where only a subset of interceptors has direct access to the target states, and the time-to-go \eqref{eq:tgohat}. Over a directed actuation graph $\mathcal{G}^a$, let each interceptor cooperatively exchange its guidance command 
    \begin{align}
        a_{M_i} =& {V_{M_i} \dot{\hat{\theta}}_i} + \left[ \frac{V_{M_i} \left(V_{M_i}^2-\hat{V}_{T_i}^2\right) \cos^2 \hat{\delta}_i}{\hat{r}_i \hat{V}_{\theta_i}} \right]\nonumber\\
        &\times
        \left( \Lambda_1 |s_i|^{1/2} \operatorname{sign}(s_i) - \nu_i(t)\right)\label{eq:am},
    \end{align}
    where $s_i=\sum _{j=1}^n[\mathcal{L}_I]_{ij} e_j$ , $ \dot{\nu}_i(t) = -\Lambda_2 \operatorname{sign}(s_i)$ for some $\Lambda_1,\Lambda_2>0$. Then, following the command \eqref{eq:am}, the interceptors establish a consensus in their time-to-go values within a finite-time $T_c$ to intercept the moving target simultaneously at a time cooperatively decided during engagement.
\end{theorem}
\begin{proof}
    We omit the detailed steps due to space constraints, but refer the reader to \cite{p6,11018241} for the derivations for cooperative salvo guidance strategies under various graph topologies with different values of consensus time.
\end{proof}
After consensus in time-to-go values is established, $s_i=0$, and hence, $a_{M_i} = {V_{M_i} \dot{\hat{\theta}}_i}$, which is a pursuit guidance term. Moreover, as $t\to T_s$, $\dot{\hat{\theta}}_i\to \dot{{\theta}}_i$, and so do all the engagement variables that are estimated. Therefore, after a transient of acceptable short duration (depending on $T_s$ and $T_c$), the interceptors attain their requisite courses to intercept the target simultaneously. Meanwhile, $s_i=0$ also implies $\dot{\hat{\delta}}_i =0$, meaning that the deviation angles get fixed in the steady-state.
    
    \section{Simulations}
    
    \begin{figure}[h!]
    \centering
    \begin{subfigure}[b]{\linewidth}
        \centering
        \includegraphics[width=.7\linewidth]{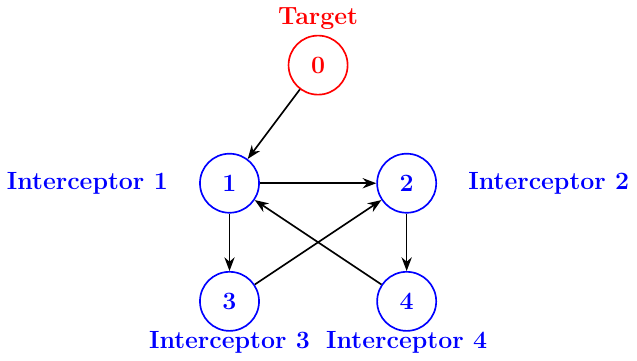}
        \caption{Sensing graph $\mathcal{G}$.}
        \label{sensing_topology}
    \end{subfigure}
    \hfill
    \begin{subfigure}[b]{\linewidth}
        \centering
        \includegraphics[width=.7\linewidth]{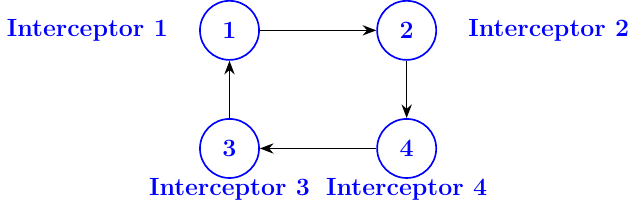}
        \caption{Actuation Graph $\mathcal{G}^a$.}
        \label{fig:actuation_graph}
    \end{subfigure}
    \caption{Interceptors' network topologies.}
    \label{fig:topologies}
\end{figure}

   Consider an engagement scenario involving one target and four interceptors (labeled I1 through I4), where only I1 is equipped with a seeker and can directly track the target, the rest of the interceptors being seeker-less. The sensing topology \( \mathcal{G} \) is illustrated in \Cref{sensing_topology}, and the actuation graph \( \mathcal{G}^a \) is shown in \Cref{fig:actuation_graph}. The target is launched from (14000,0) m, with a heading angle of $120 ^\circ$ and speed of 500 m/s. The interceptors' speeds are heterogeneous, taken as $\begin{bmatrix}
       580 & 590 & 600 & 580
   \end{bmatrix}$ m/s, whereas their initial heading angles are $\begin{bmatrix}
       15^\circ & 20^\circ & 30^\circ & 35^\circ
   \end{bmatrix}$, and  and their initial positions are $\begin{bmatrix}
       (4.5,0) & (6,0) & (7,0) & (8,0)
   \end{bmatrix} \times 10^3$m. The interceptors are assumed to operate under actuator constraints that restrict their maximum achievable lateral acceleration to 40\,g ($g$ being the acceleration due to gravity). The initial observer estimate for I1 is initialized with the true target state with an added random perturbation, while the initial estimates for the remaining interceptors are assigned arbitrarily. The observer gains are set to $k_1=0.9$, $k_2=4$, and $k_3=5$. The parameters $\alpha$ and $\beta$ are chosen as 0.93 and 1.3, respectively. The controller gains $\Lambda_1$ and $\Lambda_2$ are set to 5 and 10, respectively.
   


\begin{figure*}[h!]
    \centering
    \begin{subfigure}[t]{0.475\linewidth}
        \centering
        \includegraphics[width=\linewidth]{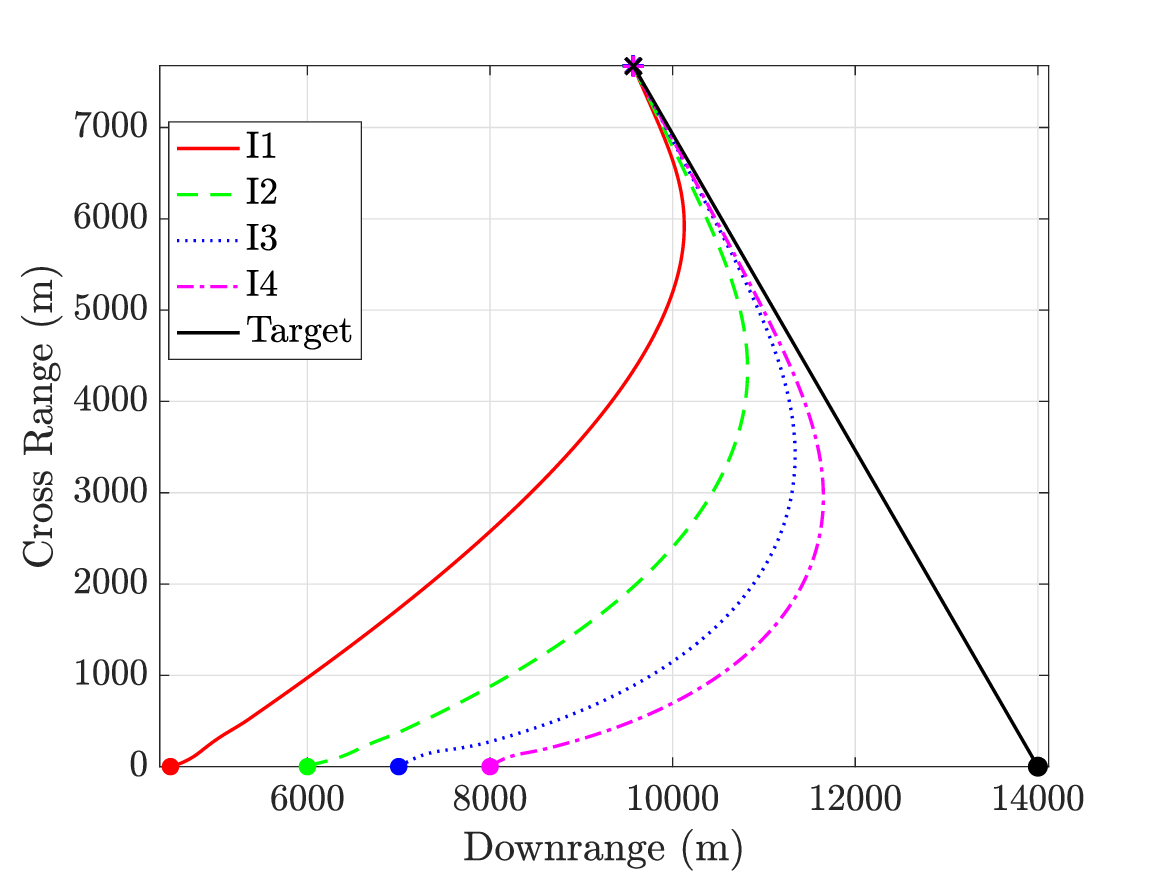}
        \caption{Trajectories.}
        \label{fig:traj1}
    \end{subfigure}
    \begin{subfigure}[t]{0.475\linewidth}
        \centering
        \includegraphics[width=\linewidth]{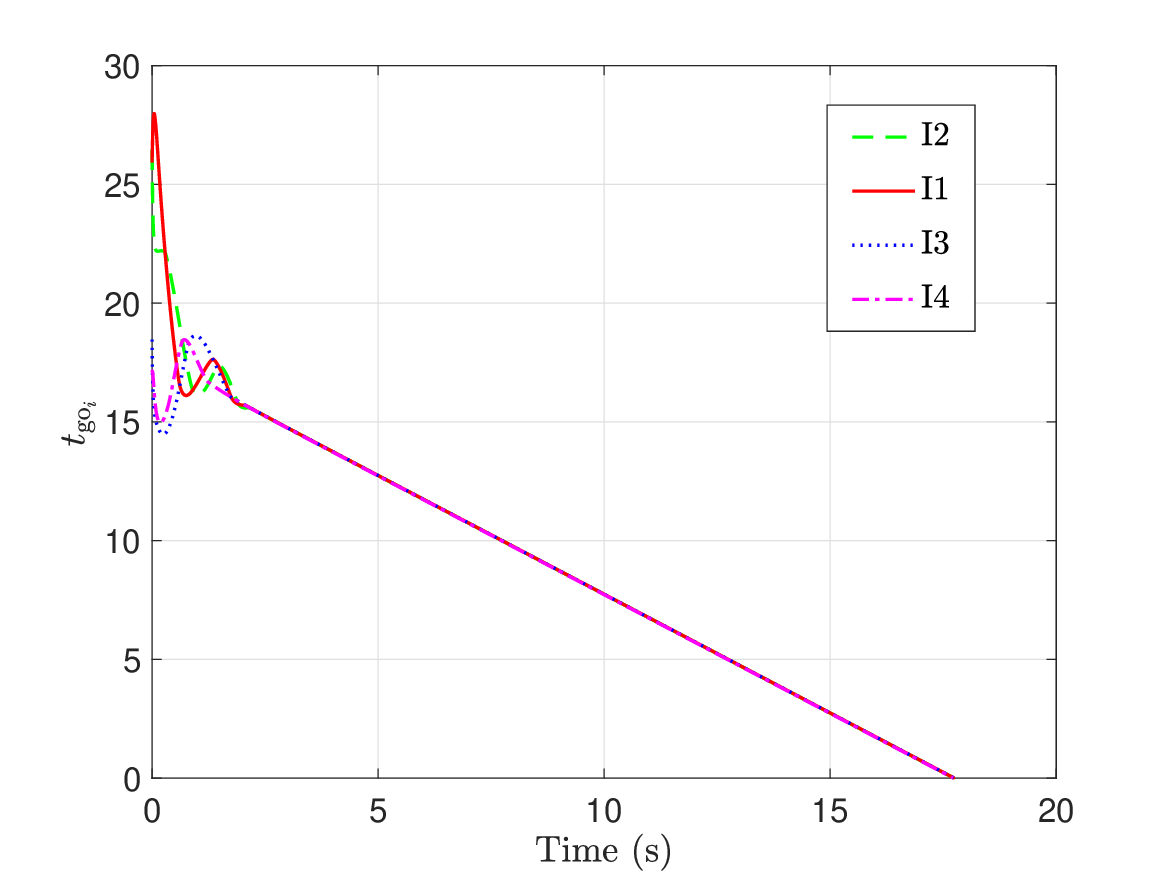}
        \caption{Time-to-go.}
        \label{fig:tgo1}
    \end{subfigure}
    \begin{subfigure}[t]{0.475\linewidth}
        \centering
        \includegraphics[width=\linewidth]{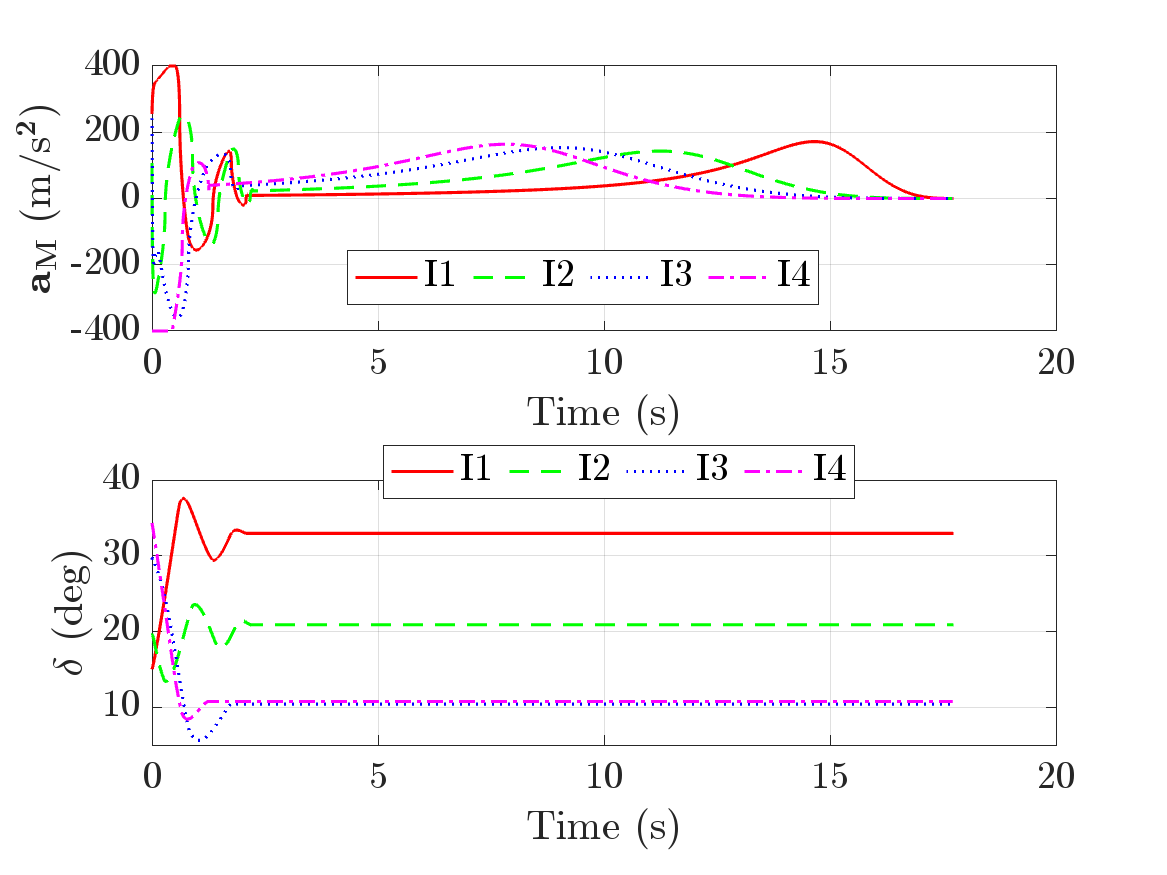}
        \caption{Lateral accelerations and deviation angles.}
        \label{fig:am1}
    \end{subfigure}
    \begin{subfigure}[t]{0.475\linewidth}
        \centering
        \includegraphics[width=\linewidth]{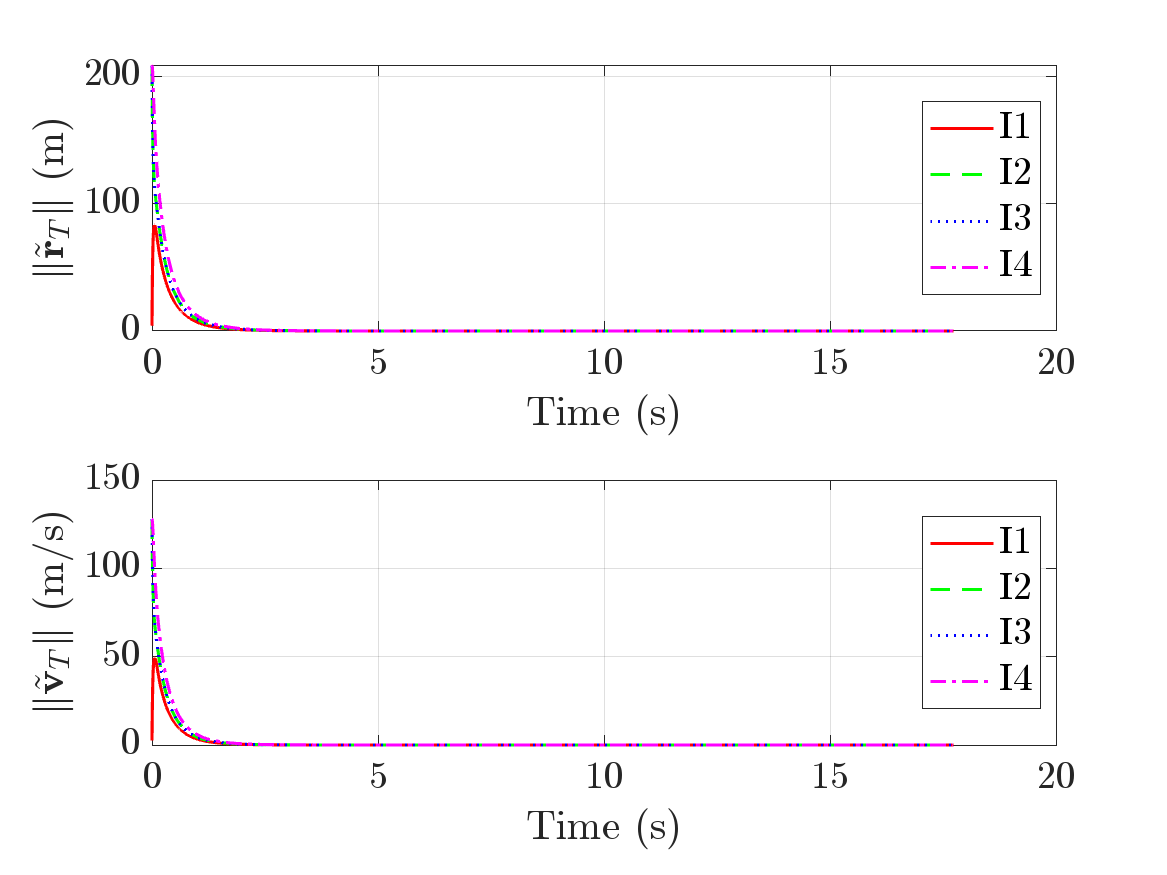}
        \caption{Observer Errors.}
        \label{fig:e1}
    \end{subfigure}
    \caption{Performance of the proposed method for a typical case.}
    \label{fig:observer_performance}
\end{figure*}

\Cref{fig:observer_performance} depicts a scenario where the interceptors achieve a cooperative simultaneous interception at 17.27 s against a moving target under incomplete information. The interceptors are launched from various locations (\Cref{fig:traj1}), where only I1 has the knowledge of the target states. The rest of the interceptors are seeker-less and rely on I1 following $\mathcal{G}$ to estimate the target states in a distributed manner. One may observe from \Cref{fig:tgo1} that, regardless of this information asymmetry, the interceptors agree on a common time-to-go value within a finite time (around 2.2 s) to attain the desired geometry leading to simultaneous interception. The lateral accelerations and the deviation angles of each interceptor are shown in \Cref{fig:am1}, where one may notice that before consensus in time-to-go, the deviation angles are changing and the lateral acceleration demands are initially higher. Once the observer error settles to zero around the same time (\Cref{fig:e1}) and the consensus in time-to-go is achieved, the deviation angles get fixed. In \Cref{fig:e1}, $\tilde{\mathbf{r}}_T = \begin{bmatrix}
    \tilde{X}_T&\tilde{Y}_T
\end{bmatrix}$ and $\tilde{\mathbf{v}}_T = \begin{bmatrix}
    \tilde{\dot{X}}_T&\tilde{\dot{Y}}_T
\end{bmatrix}$ are seen to be converging to zero rapidly.
 \begin{figure*}[h!]
    \centering
    \begin{subfigure}[t]{0.475\linewidth}
        \centering
        \includegraphics[width=\linewidth]{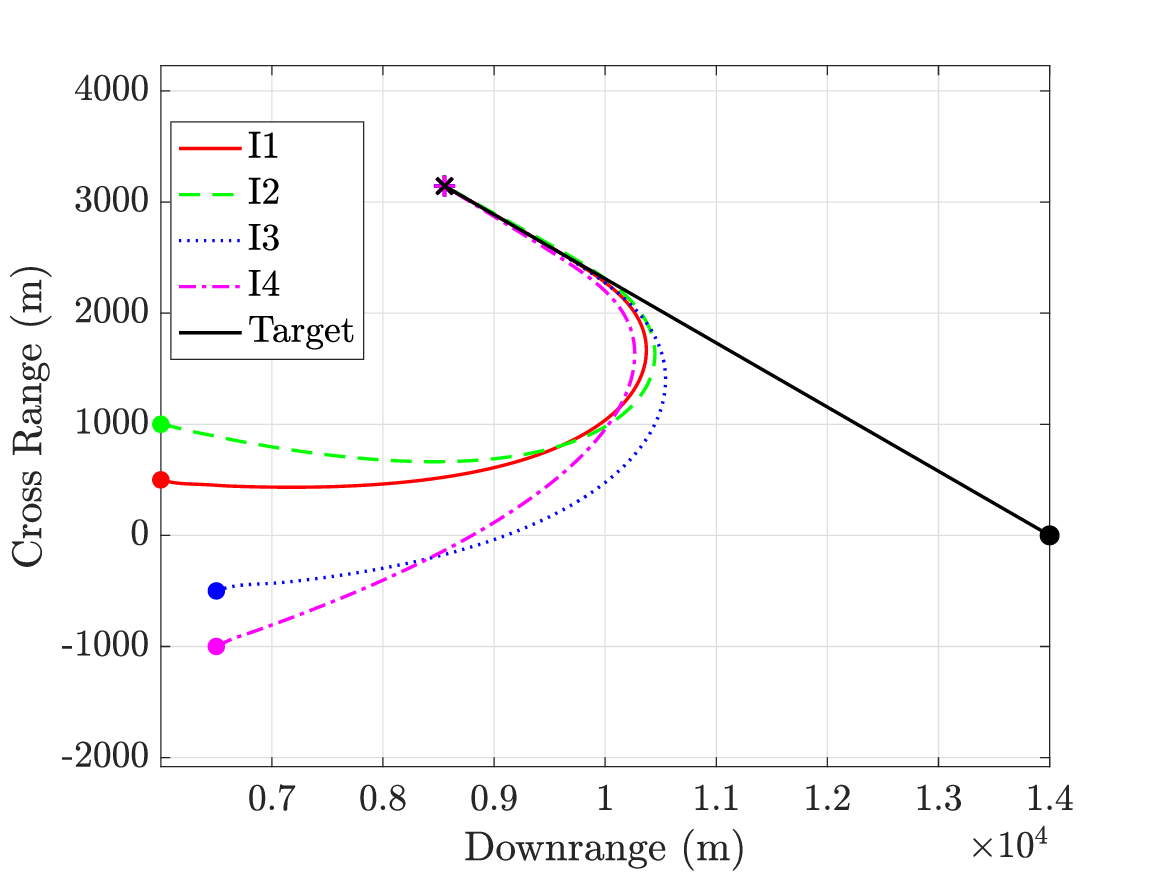}
        \caption{Trajectories.}
        \label{fig:traj2}
    \end{subfigure}
    \begin{subfigure}[t]{0.475\linewidth}
        \centering
        \includegraphics[width=\linewidth]{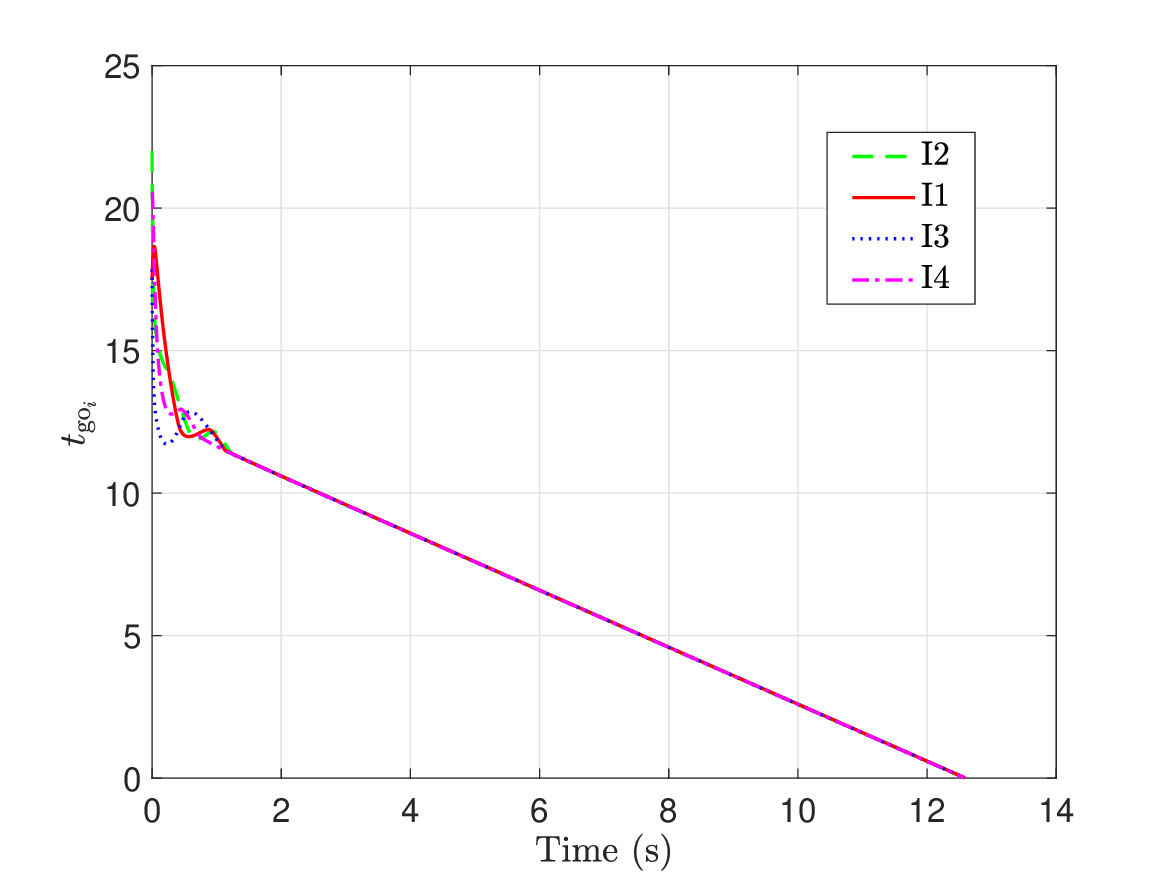}
        \caption{Time-to-go.}
        \label{fig:tgo2}
    \end{subfigure}
    \begin{subfigure}[t]{0.475\linewidth}
        \centering
        \includegraphics[width=\linewidth]{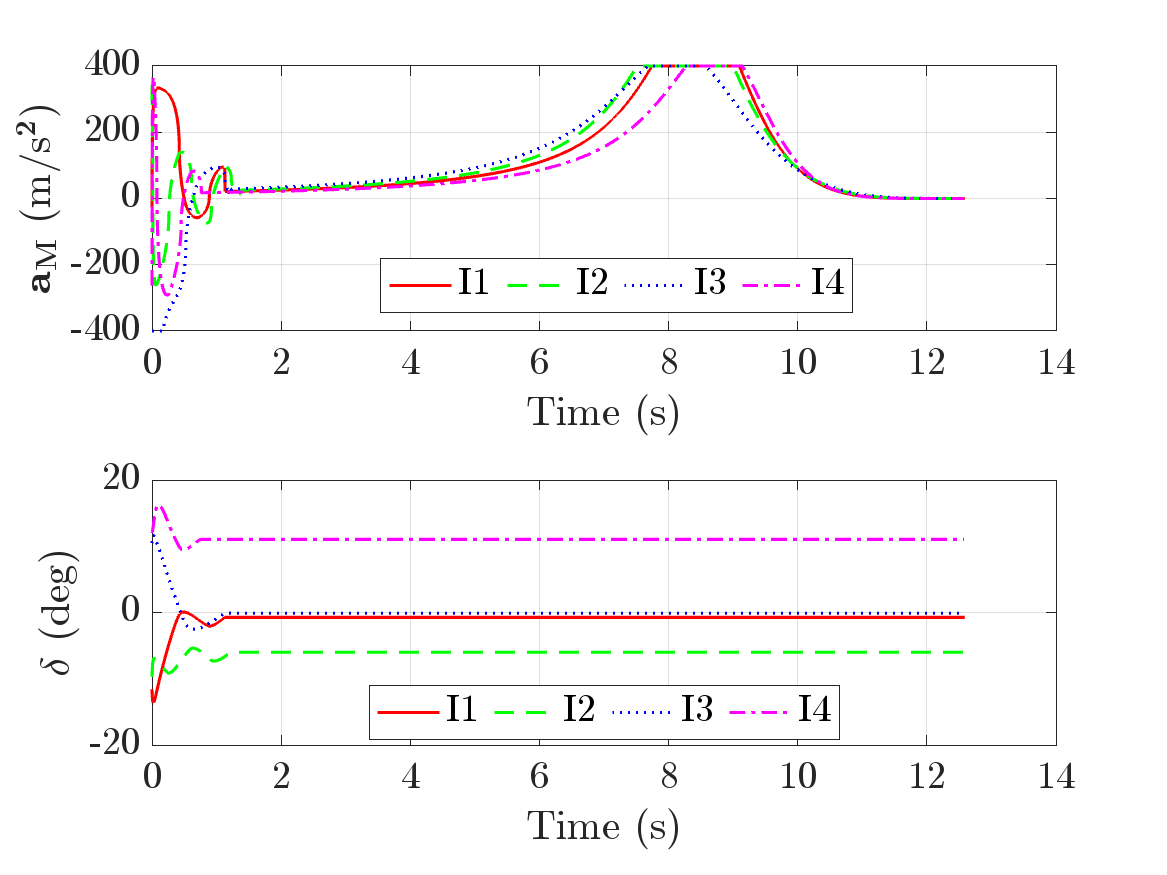}
        \caption{Lateral accelerations and deviation angles.}
        \label{fig:am2}
    \end{subfigure}
    \begin{subfigure}[t]{0.475\linewidth}
        \centering
        \includegraphics[width=\linewidth]{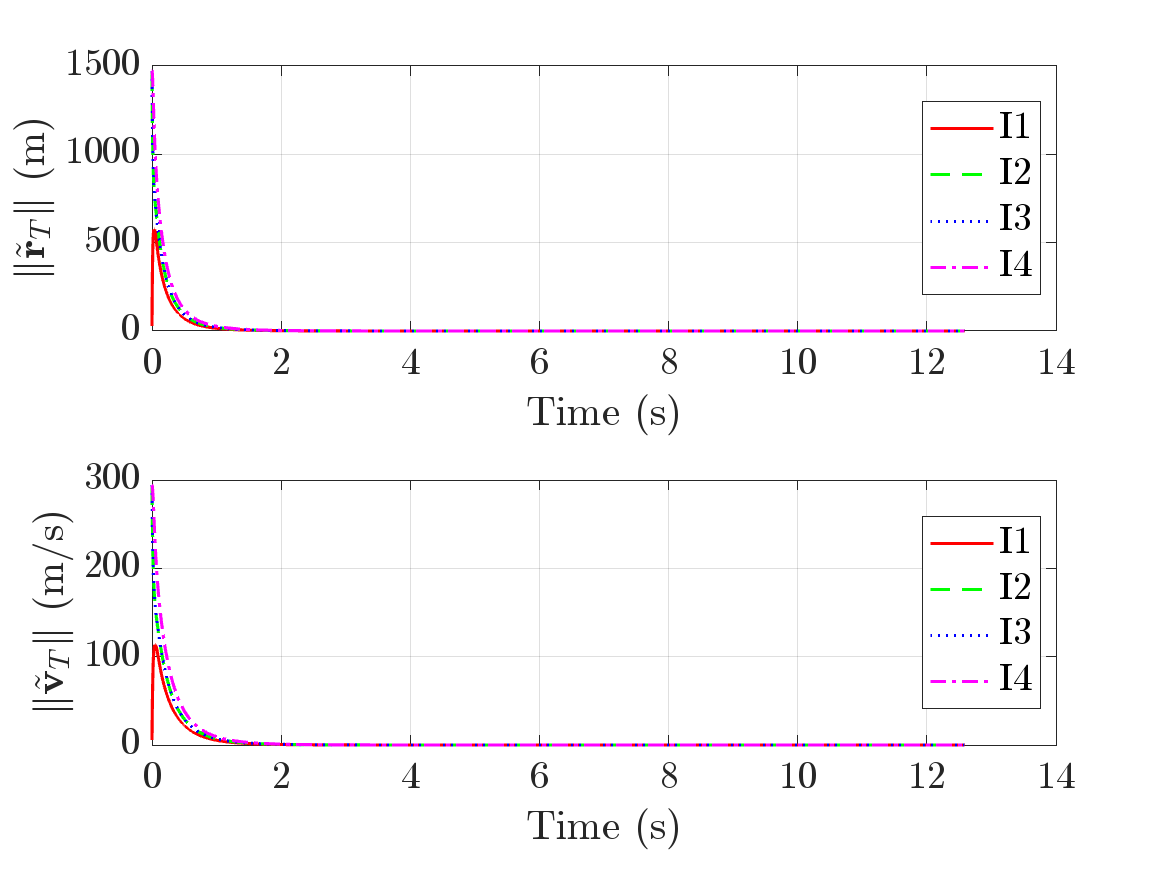}
        \caption{Observer Errors.}
        \label{fig:e2}
    \end{subfigure}
    \caption{Performance of the proposed method under a different engagement geometry.}
    \label{fig:observer_performance2}
\end{figure*}

To further assess the effectiveness of the proposed method, the result in \Cref{fig:observer_performance2} shows a scenario when the engagement geometry is changed. The target heading angle is changed to 150$^\circ$. The interceptors now have the initial heading angles as $\begin{bmatrix}
       -15^\circ & -10^\circ & 20^\circ & 25^\circ
   \end{bmatrix}$, and their initial positions are $\begin{bmatrix}
       (6,0.5) & (6,1) & (6.5,-0.5) & (6.5,-1.0)
   \end{bmatrix} \times 10^3$m. One may observe that despite changes in the initial configuration, the time of observer error convergence and the time of consensus in time-to-go do not differ significantly. However, this change in engagement geometry leads to a simultaneous interception at a different final value of around 12.57 s, which is earlier than the case shown in \Cref{fig:observer_performance}.

    \section{Conclusions and Future Work}\label{sec:conclusions}
    In this paper, we developed a cooperative guidance architecture for the time-coordinated interception of a constant-velocity, non-maneuvering target under heterogeneous sensing capabilities. By integrating a fixed-time distributed observer with an exact time-to-go formulation derived from deviated pursuit guidance, the proposed strategy enabled seeker-less interceptors to estimate target states and synchronize impact times using only local communication. A higher-order sliding mode consensus protocol ensured finite-time agreement on the interception time-to-go, avoiding the limitations of average-based or linearized methods. The proposed method shows guaranteed simultaneous interception under realistic sensing and communication constraints. Our future research will explore extensions to maneuvering targets with uncertain dynamics, time-varying communication topologies, and stochastic disturbances.
	
	\bibliographystyle{IEEEtran}
	\bibliography{references2}
\end{document}